# Master curve of boosted diffusion for ten catalytic enzymes

Ah-Young Jee,[a] Tsvi Tlusty,[a,b] and Steve Granick[a,b,c*]

[a]Center for Soft and Living Matter, Institute for Basic Science (IBS), Ulsan 44919, South Korea

[b]Departments of Physics and [c]Chemistry, Ulsan National Institute of Science and Technology (UNIST), Ulsan 44919, South Korea

**Abstract**

Molecular agitation more rapid than thermal Brownian motion is reported for cellular environments, motor proteins, synthetic molecular motors, enzymes, and common chemical reactions, yet that chemical activity couples to molecular motion contrasts with generations of accumulated knowledge about diffusion at equilibrium. To test the limits of this idea, a critical testbed is mobility of catalytically-active enzymes. Sentiment is divided about reality of enhanced enzyme diffusion with evidence for and against. Here a master curve shows that enzyme diffusion coefficient increases in proportion to the energy release rate – the product of Michaelis-Menten reaction rate and Gibbs free energy change ($\Delta G$) – with the highly satisfactory correlation coefficient of 0.97. For ten catalytic enzymes (urease, acetylcholinesterase, seven enzymes from the glucose cascade cycle, and another), our measurements span from roughly 40% enhanced diffusion coefficient at high turnover rate and negative $\Delta G$ to no enhancement at slow turnover rate and positive $\Delta G$. Moreover, two independent measures of mobility show consistency, provided that one avoids undesirable fluorescence photophysics. The master curve presented here quantifies the limits of both ideas, that enzymes display enhanced diffusion and that they do not within instrumental resolution, and has possible implications for understanding enzyme mobility in cellular environments. The striking linear dependence of $\Delta G$ for the exergonic enzymes ($\Delta G < 0$) together with the vanishing effect for endergonic enzyme ($\Delta G > 0$) are consistent with a physical picture where the mechanism boosting the diffusion is an active one, utilizing the available work from the chemical reaction.

**Significance**

Mounting evidence of boosted molecular mobility during enzyme catalysis shows that chemical activity is transduced into motion. This challenges a central tenet of chemistry and biochemistry that enzyme translational mobility is governed solely by Brownian motion, a matter of interdisciplinary interest regarding living matter as well as significant chemical technology, but previous literature is inconsistent regarding the existence of this effect in different enzyme systems. This paper surveys ten different catalytic enzymes and shows that the magnitude of enhanced diffusion scales with energy release rate, the Gibbs free energy of reaction multiplied by the Michaelis-Menten reaction rate. This master curve to predict the magnitude of boosted molecular mobility may be useful to estimate the effect in as-yet untested enzyme systems.

## Introduction

We are interested in the diffusive-looking random walks executed when chemical reactions inject energy into the system, kicking it off equilibrium. Bacteria and active colloids, for example, are propelled along trajectories that appear diffusive at long timescales, but at shorter times become more persistent and super-diffusive (1). Recent reports of "enhanced" or "boosted" diffusion of various molecules extend this notion down to the molecular scale, raising questions about the shared and distinguishing features of equilibrium and nonequilibrium, not only for protein molecular motors where the phenomenon was first identified (2) but also synthetic molecular motors (3-5), active fluctuations in the cell (6, 7) and optically trapped colloids (8), common chemical reactions (9), and enzymes (10-18). Regarding enzymes, however, experiments have not tackled the basic question as to how reaction rate and the free energy released each affect mobility. Moreover, measuring boosted enzymatic motion has proved to be challenging as it is a rapid process occurring over length scales of a few nanometers. Some studies assert that the apparent enhancement originates from experimental artifacts (19-22) as has been reviewed critically (23).

In the present work, we thus investigate ten different enzymes whose energy release rate varies over a wide range, using two independent instrumental methods. Tracing the enzymatic reaction rate, we find strikingly simple dependence of diffusivity on rate, thus quantifying the coupling of mobility and chemical activity.

## Results

**Experimental design.** Our experimental design is guided by the following considerations. To minimize chemical differences, for two of the enzymes we tune the catalytic activity by varying temperature and pH independently. To test the validity of fluorescence measurements, we compare to an independent measurement with a complementary technique, dynamic light scattering, though only fluorescence can probe the most desirably dilute concentrations. We selected enzymatic reactions whose Gibbs free energy change $\Delta G$ and the turnover rate $k_{cat}$ are either known from the literature or can be measured by us in this study. Urease was studied in several early studies reporting enhanced diffusion (11, 13-18), and is therefore included, while mindful that the hydrolysis of urea releases $CO_2$. This is potentially confounding, as it is unclear whether $CO_2$ is fully dissolved or forms bubbles, but the consistency we report with the other enzymes suggests the former. We exclude catalase (12-14) because its reaction produces visible oxygen bubbles that can generate the appearance of convection-induced enhanced diffusion for spurious reasons (24). To avoid the reported tendency of multimeric enzymes to dissociate when the substrate concentration exceeds a level roughly equal to the Michaelis constant (18), we are careful not to exceed this limit. We avoid aggregation caused by high enzyme concentration (18). Finally, a major design consideration was picking enzymes whose catalytic turnover rates vary widely, being relatively fast for some enzymes, relatively slow for others. On physical grounds, we hypothesized that if catalysis events induce enhanced diffusion, the effect should increase with turnover rate.

The ten enzymes we studied are summarized in Table I. For each, the table lists its turnover rate $k_{cat}$, Michaelis constant $k_M$, free energy of reaction $\Delta G_{rxn}$, diffusion coefficient $D_0$, measured in this laboratory in the absence of substrate, and the maximal boosted diffusion at 25 °C. Among the seven exergonic enzymes ($\Delta G < 0$), urease was selected as it is the enzyme for which enhanced diffusion was first reported (11); here we extend the $k_{cat}$ range by varying temperature and pH. For acetylcholinesterase, we also extend the range of $k_{cat}$ beyond that in the original study (15, 16) by varying temperature and pH. The other enzymes – aldolase, phosphoglucoisomerase, pyruvate kinase, hexokinase, and phosphofructokinase, were selected from the glucose cascade cycle (25). Among the three endergonic reactions ($\Delta G > 0$), alkaline phosphatase and triosephosphate isomerase were selected to repeat the measurements of an earlier study (14, 21), and phosphoglycerate kinase was selected from the glucose cascade cycle. The substrate concentrations were selected to give reaction half-lives of a few minutes.

Each of our data points represents the average of 20-30 repeated independent measurements. For many measurements, we used fluorescence correlation spectroscopy (FCS), a standard method to measure molecular diffusion in the nM regime (Methods). The principle of measurement is that fluorescence intensity fluctuates due to molecules diffusing into and out of tiny spaces whose volumes are diffraction-limited. We also used dynamic light scattering (DLS), another standard method whose principle is to quantify the photon autocorrelation function and extract from it the implied translational diffusion coefficient. FCS has the advantage of sensitivity. DLS, whose sensitivity is lower, has the advantage of no need for labeling. Pushing the limits of DLS sensitivity, we were able in some cases to almost match enzyme concentration in both experiments.

**Boosted diffusion correlates with reaction rate.** Boosted diffusion during enzyme-catalyzed chemical reactions was normalized to Brownian diffusion as $\Delta D/D_0$, where $\Delta D = D - D_0$, $D$ is enzyme diffusion coefficient during chemical reaction, and $D_0$ is the bare thermal diffusion coefficient, in the absence of substrate or when the reaction is complete. Standard Michaelis-Menten reaction kinetics give substrate concentration $c(t)$ and reaction rate, $V(t) = k_{\text{cat}} c(t)/(K_M + c(t))$. For three representative reactions plotted against time, catalysed by the enzymes phosphoglucoisomerase, acetylcholinesterase, and urease, one sees that the reaction rate $V(t)$ and the enhanced diffusion $\Delta D(t)/D_0$ decay over the same time scale (Fig.

1a) with the same fractional changes – their ratio is unity within the experimental uncertainty (Fig. 1b). Therefore, otherwise stated, $\Delta D/D_0$ ($t$) is directly proportional to $V(t)$. Fluorescence-based measurements agree quantitatively with an independent DLS measurement (Fig. 1b).

**Experimental validation of FCS.** Before presenting our main results, we examine the scope and limitations of this treatment, and its relation to previous empirical and theoretical discussions of the enhanced enzyme diffusion problem. To check the reliability of our FCS data, we performed direct tests of fluorophore photostability. To find a fluorophore whose photostability allows diffusion to be deduced from the FCS intensity-intensity autocorrelation curve, we screened candidate dyes and selected Atto 488 based on observing that its fluorescence lifetime decay, when bound to the enzymes of interest, was the same in the presence and absence of substrate under the respective buffer conditions of each enzyme reaction, and likewise, in three cases studied explicitly, the same in the presence of product but no reactant. Illustrating this, fluorescence intensity decay on the nanosecond timescale, measured using time-correlated single photon counting, is plotted for pyruvate kinase, hexokinase and acetylcholinesterase (enzymes 3, 4, and 6, respectively) in Fig. 1C. All fluorescence decay curves were nonexponential as is typically found since there exist multiple mechanisms of excited state deactivation (26). We fitted them by double exponential functions. Because both timescales affected the overall fluorescence lifetime, the average fluorescence lifetime $<\tau>$ (with $<\tau> = A_1\tau_1 + A_2\tau_2$, $A_1 + A_2 = 1$) was used for further analysis. For all ten enzymes, the average fluorescence lifetimes at room temperature for Atto 488, with substrate present or absent, are shown in Fig. 1D. In contrast, some of the enzymes labeled with the oft-employed dyes Alexa Fluor 488 and Cy3 (Fig. 1E) presented significant lifetime decrease during the respective chemical reactions that they catalyze. Thus, although these latter two dyes are sometimes regarded as a gold standard, they were not employed in our subsequent measurements. These precautions were necessary to avoid the known corrupting influence that photophysical changes, such as photobleaching and reversible quenching, can have on fluorescence intensity-intensity fluctuations, shifting the autocorrelation curve to shorter times for this spurious reason (22, 27). Inconsistent findings in the enzyme community can be attributed in part to using different fluorescent probes. In the following, we report on the consistency of FCS and DLS measurements for various temperature and pH conditions.

**Master curve of boosted diffusion for ten enzymes.** The seven exergonic ($\Delta G < 0$) enzymes in Table I exhibit boosted diffusion with a magnitude proportional to the reaction rate $V$ with a correlation coefficient 0.650 (Fig. 2A). Multiplying $V$ by $-\Delta G$ we find the energy release rate. We see that the boosted diffusion $\Delta D/D_0$ is directly proportional to the energy release rate with a correlation coefficient of 0.986 for measurements at 25°C (Fig. 2B) and 0.970 when measurements at all temperatures and pH are included (Fig. 2C). In contrast, these data do not correlate significantly with the enthalpy change $\Delta H$ (Fig. 3A) nor with the enthalpy release rate $V(-\Delta H)$ (Fig. 3B).

One notable case is the enzyme alkaline phosphate, which catalyzes a highly exothermic reaction with $\Delta H = -43.5$ kJmol$^{-1}$ yet positive $\Delta G = +30.2$ (28) or $+61.5$ (29) kJmol$^{-1}$ at a high turnover rate. This shows no detected enhanced diffusion in our study, in agreement with a recent earlier study (21). Therefore, we consider it more meaningful physically to seek correlation with $\Delta G$, the maximum available work that can be extracted from the reaction (15, 16). Moreover, heat release would dissipate too rapidly to induce enhanced diffusion (16, 30). At these ultralow enzyme concentrations (nM), we found no dependence on enzyme concentration; the master curve in Figs. 2B-C appears to be a single-enzyme property, not a collective effect. The striking linear dependence of $\Delta G$ for the exergonic enzymes ($\Delta G < 0$) together with the vanishing effect for endergonic enzyme ($\Delta G > 0$) are consistent with a physical picture where the mechanism boosting the diffusion is an active one, utilizing the available work from the chemical reaction.

**Temperature and pH dependence of boosted diffusion.** Our dataset includes experiments in which we tuned temperature and pH in order to compare the same enzyme at different $k_{cat}$. Because enzyme functional groups at and near active sites have a wide range of pK$_a$, changes of temperature and pH alter their charge makeup and thus modulate their catalytic efficiency and turnover rate. To measure enzyme activity under each of these conditions, standard enzyme assays were performed. The values of $k_{cat}$ were inferred for urease and acetylcholinesterase from standard Lineweaver–Burk plots with excellent fits of Michaelis-Menten parameters (Fig. 4A-B). For both enzymes, the turnover rate increases monotonically with temperature (Fig. 4C). With pH varied at room temperature (Fig. 4D), it shows a maximum at pH = 7. The limits of temperature and pH that we studied were set by enzyme stability. It is reassuring that the Gibbs free energy ($\Delta G$) values implied by the temperature data, -21.05 and -16.16 kJ/mol for urease and acetylcholinesterase respectively, are consistent with known values (28, 31). To orient the reader, a typical small protein with diffusion coefficient $D \approx 50$ μm$^2$-s$^{-1}$ and $k_{cat} \approx 10^3$ s$^{-1}$ diffuses at equilibrium the root-mean-square distance $\approx 100$ nm in 1 ms, the average time between its catalytic events.

**Comparison of FCS with DLS at various temperature and pH.** Dynamic light scattering (DLS) validated consistency with fluorescence-based FCS measurements. Our DLS tests (Methods) were performed on enzymes labeled in the same manner as for the FCS experiments, with the same buffer conditions and the same temperature. Working within the constraint that DLS is less sensitive than FCS and therefore demanded higher enzyme concentrations, we employed DLS enzyme concentrations as close as possible to those used for FCS. Fig. 4E-F shows quantitative agreement in measurements of urease (#7) and acetylcholinesterase (#6), both in measurements at pH = 7 and different temperatures (Fig. 4E) and at 25 ºC and different pH (Fig. 4F). Diffusion coefficients measured by these independent methods agree within the experimental uncertainty.

**Discussion**

We employ the term "boosted diffusion" to emphasize that this phenomenon is associated with chemical activity such that released chemical energy generates persistent motion (9, 15, 16) and thereby increases the mobility. In this view, endergonic reactions ($\Delta G > 0$) are predicted to show no boosted mobility, as observed, while the boosts of the exergonic enzymes ($\Delta G < 0$) collapse on one master curve because they have similar timescales of reorientation and boost from chemical reaction (16). While enhanced diffusion is observable only for high turnover enzymes, it may in principle occur also in slower enzymes where it is masked by dominant Brownian motion.

The master curve we have presented is broadly consistent with that from a parallel study of non-enzymatic chemical reactions (9, 16) but that study showed considerably more scatter, probably because those reactions possess more complex intermediate states, in a variety of solvents, whereas the present reactions lack complex intermediates and the solvent is always aqueous solution. Theoretically, different underlying mechanisms have been proposed such as conformational changes (30, 32, 33) cross-diffusion (34, 35), exothermicity (14), momentum exchange (36), and solvation shifts owing to electronic rearrangements during chemical reactions (9). In short, the theoretical community is aware of the enhanced diffusion problem and is actively working to come up with an underlying physical mechanism – a difficult task since one needs to link normal liquid dynamics to the quantum chemistry at the angstrom regime of enzymatic reactions.

This theoretical challenge is within a new theme of "active matter" research (1) and the view that the conversion of chemical to kinetic energy by impulsive enzymes is a force in mechanobiology (37). The empirical correlations presented here are quite independent of their theoretical specific mechanistic origin and may serve as a phenomenological guide for further investigation.

The mobility we study here differs, we believe, from the "micro-motor" situation in which catalytically-active enzymes, urease in many instances, were attached chemically to colloids or vesicles and their enhanced mobility was observed when substrate was added. (38-42) Micro-motors are driven by a concentration gradient of reaction products near the surfaces of colloidal beads, the phenomenon known as diffusiophoresis. (43-45) Diffusiophoresis is not believed to contribute to the situations considered here as the nm size of enzymes and their nm concentrations are too small to establish the concentration gradients that underpin diffusiophoresis of colloids and vesicles.

Physically, the magnitudes of $\Delta D/D_0$ that we observe and their proportionality to reaction rate suggest that energy released by chemical reactions transiently propel enzymes at the nanoscale during catalytic events. Accompanied by random thermal reorientation and translation, these boosts produce random yet persistent motion over distances vastly exceeding molecule dimensions, as we analyzed elsewhere for two specific enzymes (15, 16), but the magnitude of diffusion enhancement that can be extracted from a given amount of free energy released turns out to be sensitive to parameters not yet known from direct experiment (16, 23). To elucidate quantitatively the nature of boosts stimulated by chemical reactions comprises a research challenge for future work, but a basic assumption of this proposed scenario is that a significant part of the released free energy $-\Delta G$ is transduced into persistent motion of the enzyme along distances of a few nanometers and durations of a few microseconds (16). As the enzyme is thrust along this stochastic wormlike trajectory, viscous forces gradually dissipate the released free energy into smaller and faster thermalized degrees-of-freedom of solvent molecules. We recognize that the proposed physical scenario is speculative at this stage of theoretical understanding. It is quite different from the standard textbook view of chemical reactions in which energy released by a chemical reaction is transduced into thermal motion of the solvent directly as heat.

This common mode of enhanced diffusion in living systems (46) thus generalizes pleasingly to enzyme macromolecules when they likewise consume energy by catalyzing chemical reactions. The master curve introduced here may have functional implications for how reaction rate and substrate availability regulate enzyme diffusion in cellular environments where unambiguous measurement of single-enzyme diffusion is presently not feasible.

**Acknowledgements:** This work was supported by the taxpayers of South Korea through the Institute for Basic Science, project code IBS-R020-D1.

**Table 1. Enzyme specifications.** This table lists code number to identify each enzyme and its turnover number $k_{cat}$, Michaelis-Menten constant $K_M$, Gibbs free energy of reaction $\Delta G$, diffusion coefficient measured in the absence of substrate $D_0$, and relative enhanced diffusion $\Delta D/D_0$ measured at the earliest measurement times (corresponding references in brackets).

| Code | Enzyme | $k_{cat}$ (s$^{-1}$) | $K_M$ (mM) | $\Delta G$ (kJ/mol) | $D_0$ (μm$^2$/s) | $\Delta D/D_0$ (25 °C) |
|---|---|---|---|---|---|---|
| 1 | fructose bisphosphate aldolase | 5-42 (18-20, 47) | 0.12 | -1.3 (48) | 55 ± 2.4 | 0.01 |
| 2 | Phosphofructokinase | 150 (49) | 0.15 (49) | -26 (50) | 53 ± 3.2 | 0.03 |
| 3 | pyruvate kinase | 232 (51) | 0.1 (51) | -33.4 (50) | 65 ± 3.3 | 0.037 |
| 4 | hexokinase | 250 (18) | 0.04 | -33.5 (48) | 64 ± 2.6 | 0.04 |
| 5 | phosphoglucoisomerase | 3330 (52) | 1.5 | -2.92 (31) | 62 ± 2.4 | 0.1 |
| 6 | acetylcholinesterase | 14000 (18) measured here | 0.5 (18) | -17.6 (53) | 45 ± 1.8 | 0.18 |
| 7 | urease | 17000 measured here; 2000~45000 (11, 18, 54) | 3 | -21.5 (55) | 39 ± 1.2 | 0.24 |
| 8 | phosphoglycerate kinase | 685 (56) | 0.27 | +1.3 (56) +90 (57) +20.9 (58) | 65 ± 3.1 | 0 |
| 9 | triosephosphate isomerase | 13000 (14) | 1.8 (14) | +2.5 (48) +47.3 (59) | 61 ± 2.8 | 0 |
| 10 | alkaline phosphatase | 14000 (14) 95 (29) | 1.3 0.0003 | +30.2 (53) +61.5 (29) | 54 ± 2.0 | 0 |

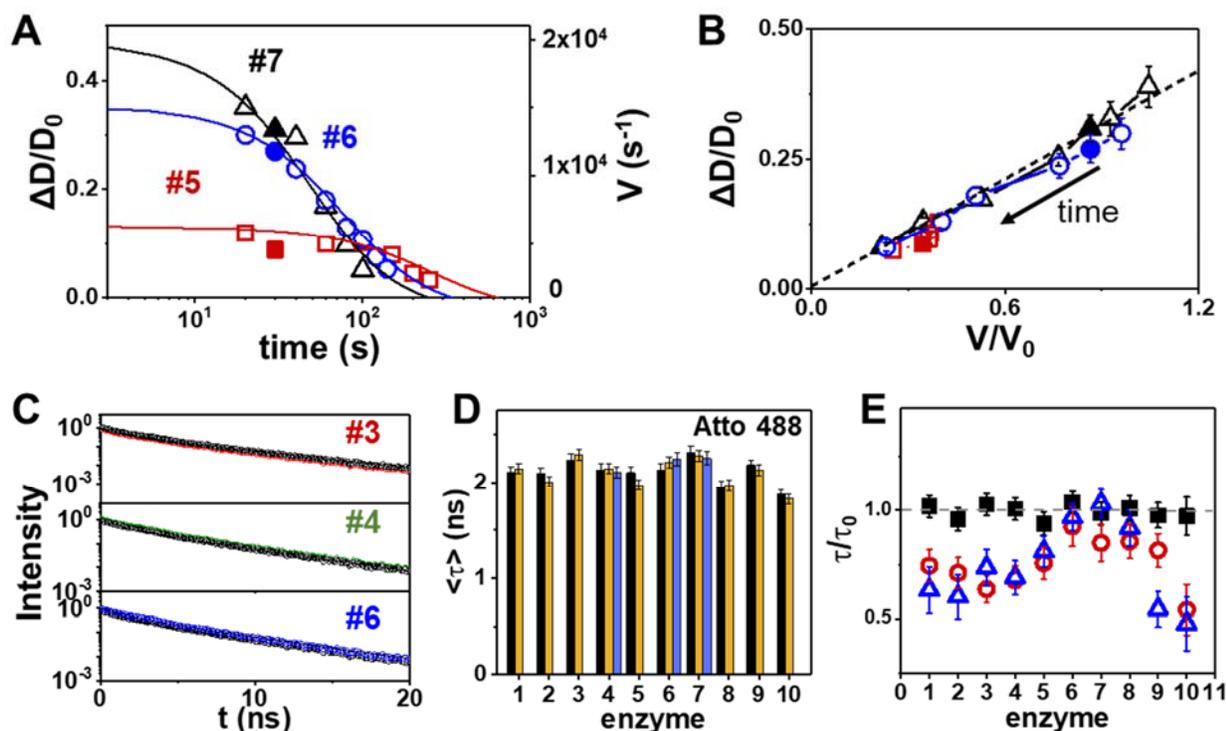

Fig. 1. **(A-B)** Boosted enzyme diffusion $\varDelta D/D_0$ scales with reaction rate. **(A)** $\varDelta D/D_0$ (left ordinate) and reaction rate $V$ (right ordinate) plotted against time (log scale) for enzymes #5 (squares), #6 (circles) and #7 (triangles) from Table 1, measured by FCS (empty symbols) and DLS (filled symbols) at 25 ºC. Solid lines are calculated from $k_{cat}$ using first-order reaction kinetics. **(B)** Same data replotted as $\varDelta D/D_0$ as a function of $V/V_0$ with the same symbols. Dashed line shows the linear relation. **(C-E)** Fluorescence lifetime decay, using Atto 488 dye, is constant across all 10 enzyme systems under the respective buffer conditions of each enzyme reaction. **(C)** Raw data illustrating fluorescence lifetime decay of enzymes #3, #4 and #6 in the presence (colored symbols) and absence of substrate (black symbols). **(D)** Bar graphs compare the average fluorescence lifetime $<\tau>$ for all 10 enzymes, in the presence (yellow bars) and absence of substrate for Atto 488 dye, and for three enzymes in the presence of product but not reactant (blue bars). **(E)** Ratio of average fluorescence lifetime $<\tau>$ with and without substrate for all 10 enzymes for Atto 488 (black squares), Alexa Fluor 488 (red circles) and Cy3 (blue triangles).

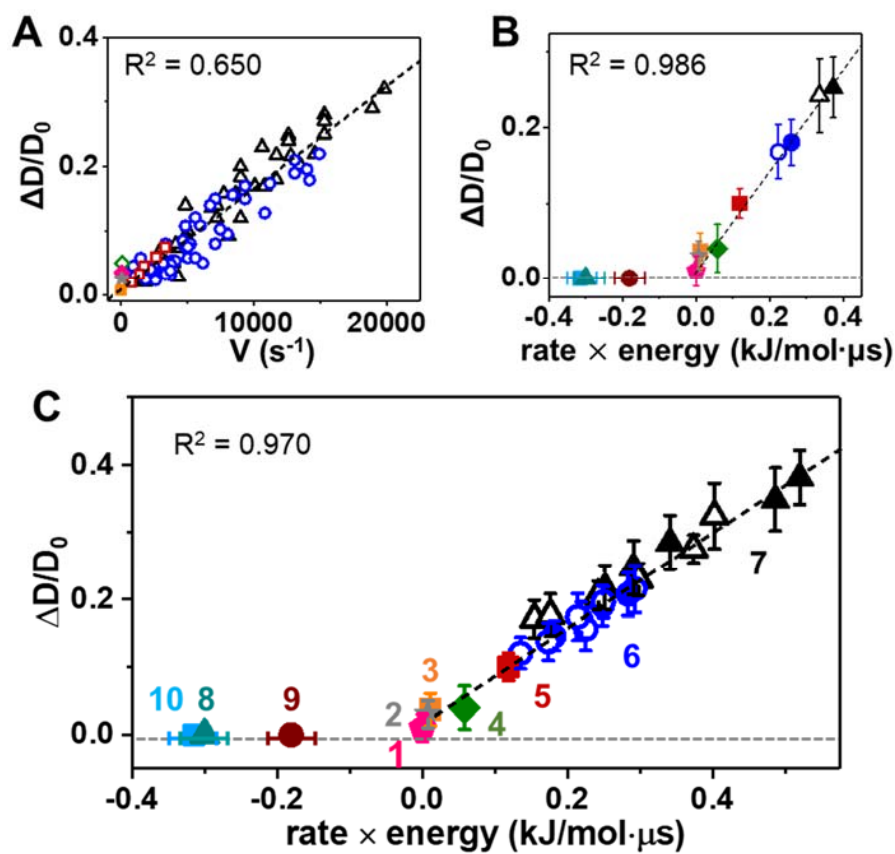

Fig. 2. Master curves. **(A)** $\Delta D/D_0$ plotted against the reaction rate $V$ for the seven exergonic enzymes ($\Delta G < 0$, #1 to #7 in Table 1), symbols as in (C). **(B)** For measurements at 25 °C, $\Delta D/D_0$ plotted against the product of $V$ and $(-\Delta G)$, the free energy release rate. For the three endergonic enzymes ($\Delta G > 0$), $\Delta D/D_0 = 0$. Symbols as in (C). **(C)** $\Delta D/D_0$ plotted against $V(-\Delta G)$, the free energy release rate. Horizontal error bars for enzyme phosphoglycerate kinase (#8), triosephosphate isomerase (#9), and alkaline phosphate (#10) reflect the range of reported $k_{cat}$ (Table I). Symbols are identified by enzyme code in Table 1. Symbols are identified by enzyme code number in Table I. Data are taken at 25 °C and constant pH except for auxiliary measurements with temperature and pH varied for enzymes #6 and #7 with details given in Fig. 4. The units of energy release, here in kJ/mol·μs, are equivalent to $k_BT/\mu s$ per molecule ($k_BT = 2.479$ kJ/mol).

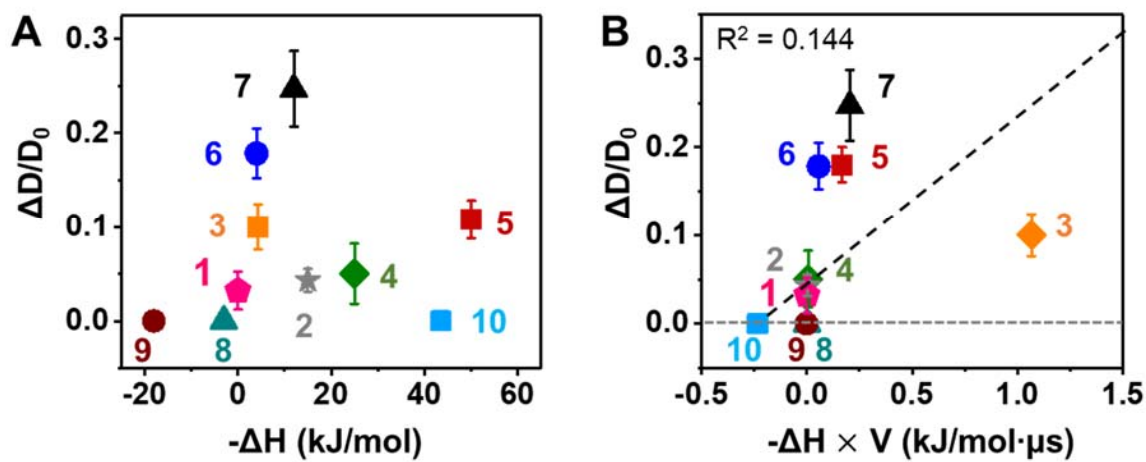

Fig. 3. Boosted diffusion shows minimal correlation with enthalpy of reaction. (**A**) *ΔD/D₀* plotted against *ΔH*. (**B**) *ΔD/D₀* plotted against the enthalpy release rate *V(-ΔH)*. The dotted line, the best linear fit, has $R^2$ = 0.15, significantly weaker than $R^2$=0.970 presented in Fig. 2D.

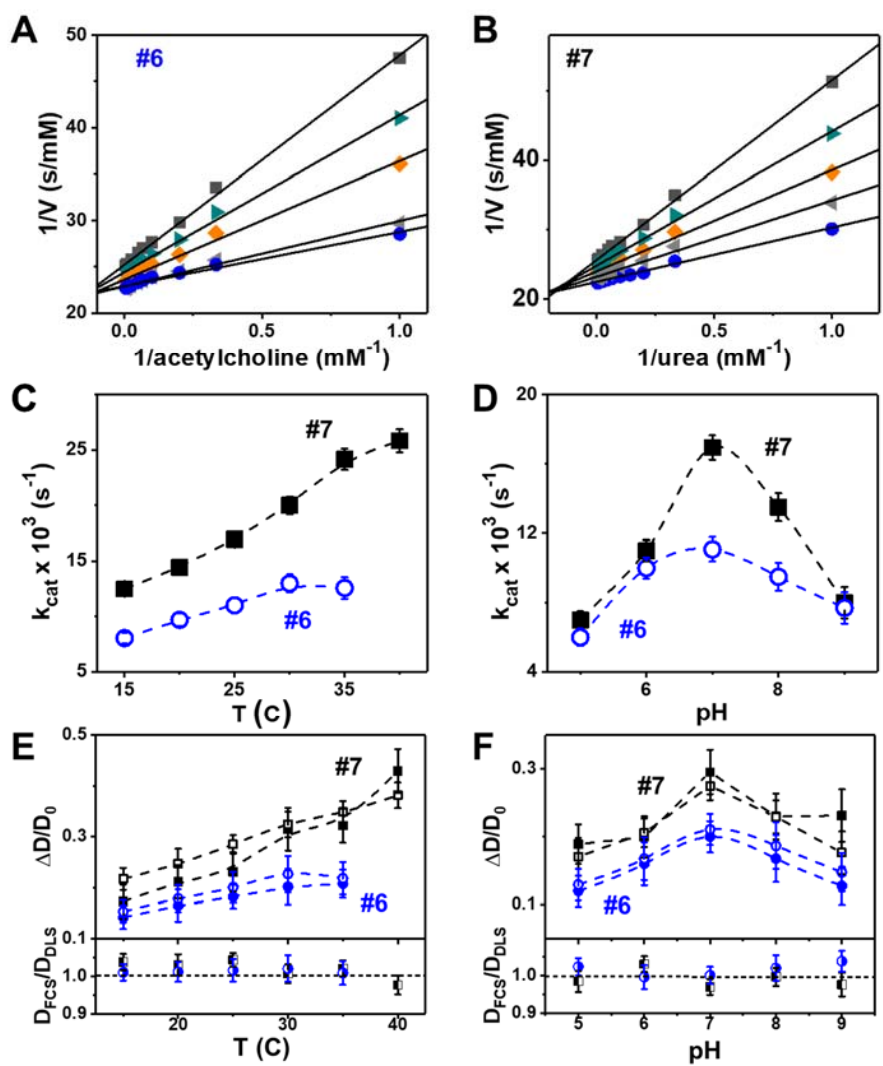

Fig. 4. Adjusting $k_{cat}$ according to temperature and pH with diffusion measured using independent methods. **(A-B)** Lineweaver-Burk plots for acetylcholinesterase (A) and urease (B) to deduce $k_{cat}$. Inverse reaction rate is plotted against inverse substrate concentration. **(C-D)** The deduced $k_{cat}$ plotted against temperature (°C) and pH (D) for acetylcholinesterase (circles) and urease (squares). **(E-F)** Dynamic light scattering validates FCS measurements. **(E)** $\Delta D/D_0$ measured by DLS (filled symbols) and FCS (open symbols) plotted against temperature for the enzymes acetylcholinesterase #6 (blue circles) and urease #7 (black squares). Lower panel shows the ratio $D_{FCS}/D_{DLS}$ at each temperature. **(F)** $\Delta D/D_0$ plotted against pH for enzymes #6 (blue circles) and #7 (black squares). Lower panel shows $D_{FCS}/D_{DLS}$ at each pH.

**Methods**

Enzyme samples and dye labeling. Urease (type C-3) from jack bean, purchased from Sigma, was labeled at the cysteine residue with Dylight 488 maleimide dye by a protocol involving 150 mM phosphate buffer (pH 7.2) with added 2 μM urease and 40 μM fluorescent dye solution, stirred for 6 h at room temperature. Acetylcholinesterase from Electrophorus electricus (electric eel), purchased from Sigma-Aldrich, was labeled at its amine residue by Dylight 488-NHS (N-hydroxysuccinimide) dye using a protocol in which 30 μM dye solution and 1 μM enzyme were added to a mixture of 80% phosphate buffer solution (PBS) and 20% dimethyl sulfoxide (DMSO) before 6 h of stirring at room temperature. Finally, the dye-labeled enzymes were purified by removing the free dye by membrane dialysis (Amicon ultra-4 centrifugal filter; Millipore). Hexokinase I from Saccharomyces cerevisiae was purchased from Sigma-Aldrich and was labeled with Alexa fluor 488 labeling kit (Invitrogen) using a protein fluorescence labeling kit (Invitrogen). Fructose bisphosphate aldolase from rabbit muscle, pyruvate kinase (type III), phosphoglucose isomerase (type III) from baker's yeast, fructose-6-phosphate kinase (type VII) from Bacillus stearothermophilus, 3-phosphoglyceric phosphokinase from baker's yeast, triosephosphate Isomerase from rabbit muscle, and alkaline phosphatase from Escherichia coli were purchased from Sigma-Aldrich and were labeled with Alexa fluor 488 labeling kit (Invitrogen), Atto 488 protein labeling kit (Sigma), and Cy3 Protein labeling kit (Sigma).

For fluorescence labeling, a new, freshly-opened bottle of enzyme was used each time. Newly-labeled enzymes were used within at most 2-3 days. Enzyme solutions were stored at 4 °C.

During dye labeling, reaction times were selected to give at most one dye per enzyme on average as determined from comparing UV-VIS absorbance measurements of the enzyme and dye-labeled enzyme. Consistent $D_0$ measured using DLS with and without dye-labeling showed that labeling with a single dye molecule (<1 nm) produced no detectable change of $D_0$.

Enzyme substrates. Fructose-1,6-bisphosphate (aldolase reaction), fructose-6-phosphate (phosphofructokinase reaction), phosphoenolpyruvate (pyruvate kinase reaction), glucose (hexokinase reaction), glucose-6-phosphate (phosphoglucose isomerase reaction), acetylcholine (acetylcholinesterase reaction), urea (urease reaction), p-nitrophenylphosphate (alkaline phosphatase reaction), glycerol 3-phosphate dehydrogenase (triosephosphate isomerase reaction), 3-phosphoglycerate (phosphoglucerate kinase reaction).

Experimental. Bearing in mind that vortex mixing can denaturate proteins, the solutions were mixed by gentle pipetting. In our reported values of $\Delta D/D_0$ as a function of temperature and pH, $D_0$ was measured separately without substrate at each of these conditions. Measurements by FCS began 30 s after mixing. FCS measurements lasting 30 s were repeated 20-30 times. First an enzyme preparation was divided into 10 aliquots and each aliquot was measured independently in repeat experiments each time with fresh substrate. During the following day, a new enzyme preparation was made and this procedure was repeated. In some cases, this procedure was repeated during a third day. DLS measurements were made immediately after 5 s mixing. For DLS, 30 nM Atto 488 labeled enzyme was added to 50 mM substrate and measurements were made for 50 s, which is the initial half-reaction time. The enzyme concentration used for DLS was selected to be the lowest at which photon autocorrelation curves of sufficient quality were achieved. Error bars show the standard deviation.

Enzyme activity assays. The urease and acetylcholinesterase assays were performed using the urease activity kit (MAK120, Sigma Aldrich) and acetylcholinesterase activity kit (MAK119, Sigma Aldrich) as reported in the manufacturer's instructions. Activity of other enzymes listed in Table I was taken from references cited in the Table.

Fluorescence correlation spectroscopy. In preparing for FCS measurement, dye-labeled enzyme was mixed with substrate in the appropriate aqueous buffer. When studying urea (Sigma), 2 nM dye-labeled urease was added at room temperature (100 mM PBS buffer (pH 7.2)). When studying acetylcholinesterase (AChE), 2 nM dye-labeled AChE was added in acetylthiocholine (Sigma) at room temperature (100 mM PBS buffer (pH 7)). When studying hexokinase, 2 nM of dye-labeled hexokinase I was added to a medium containing 50 mM Tris:HCl pH 7.5, 0.5 mM $MgCl_2$, 0.12 mM ATP, 0.1 mM NAD(P)+, and 0.03 mM of glucose as desired 50 mM Tris-HCl (pH 8.5). When studying aldolase, 2 nM of dye-labeled aldolase was added to 50 mM HEPES buffer (pH 7.4) with 0.1 mM of fructose-1,6-bisphosphate were mixed. The pyruvate kinase (PK) reaction was carried out by adding 2 nM PK in a reaction buffer containing 50 mM potassium phosphate, pH 7.0, 30 mM potassium chloride, 1 mM $MgCl_2$, 0.1 mM ADP, 0.1 mM P-enolpyruvate, 1 mM ATP, and 0.1 mM pyruvate. When studying phosphoglucoisomerase (PG), 2 nM of PG was added to 1 mM glucose-6-phosphate dissolved in 20 mM Tris-HCl buffer, pH 7.7. When studying phosphofructokinase (PF), 1 nM PF was added to 0.1 mM fructose-6-phosphate, ATP 0.5 mM, NADH 0.2 mM, and 20 mM Tris-HC1 buffer pH 7.9. For alkaline phosphatase reaction, enzyme dispersed in diethanolamine 2 M pH 9.8, 1 mM $MgCl_2$ and 20 mM $ZnCl_2$, with 1 mM of p-nitrophenylphosphate (Sigma). triosephosphate isomerase were performed in triethanolamine 100 mM pH 7.9, in the presence of 0.125 mM NADH, 1 mM of glycerol 3-phosphate dehydrogenase (Sigma) with 2 nM enzyme. For phosphoglycerate kinase reaction, 2 nM dye labeled phosphoglycerate kinase dispersed in 20 mM Tris-HCl pH 7.5, 5 mM $MgCl_2$, 5 mM ATP, 0.2 mM NADH, and 0.2 mM of 3-phosphoglycerate.

All fluorescence measurements were made in the presence of standard photobleaching agent: Stock agent made by dissolve trolox (2 mM), cyclooctatetraene (1 mM), and nitrobenzyl alcohol (1.5 mM) (Sigma-Aldrich) in 1ml DMSO and added to each reaction to a final concentration of 1/10 dilution.(60, 61)

FCS measurements were performed in an inverted microscope using a Leica TCS SP8X, using a 100× oil immersion objective lens with numerical aperture N.A. = 1.4 and pinhole size equal to 1 airy unit. Emitted fluorescence was collected using an avalanche photodiode (APD) (Micro Photon Devices; PicoQuant) through a 500- to 550-nm bandpass filter. The excitation power was controlled up to 20 μW. The APD signal was recorded using a time-correlated single–photon-counting (TCSPC) detection unit (Picoharp 300; PicoQuant).

To begin, the samples – substrate solutions (hundreds of mM) and a relatively high (50 nM) concentration of dye-labeled enzyme -- were equilibrated in a water bath at the desired temperature for about 10 min. Temperature during FCS measurement was controlled at the sample stage and the objective lens. The enzyme solution was loaded into a Nunc 1 coverslip 8-chamber slide (Lab-Tek Chambered Coverglass, Thermo Scientific) and mixed with a small aliquot of substrate solution to give an enzyme concentration below 2 nM with the desired sub-mM substrate concentration at a total volume of 300 μL. FCS measurement began about 30 seconds after mixing.

The FCS setup was freshly aligned for each channel on the day of use. For this purpose, a solution should be used of dye with known diffusion coefficient (1-10 nM), with similar excitation and emission wavelengths as the sample of interest; we selected solutions of Alexa 488 (D = 435 $\mu m^2/s$). The calibration chamber was equivalent to the one used to measure samples. For calibration and subsequent experiments, first we measured scattering from the cover slip glass surface in order to determine its location, then we focused 10 μm into the solution. Our control experiments showed that diffusion coefficient of reference and enzyme sampled did not depend on focus position in the range 5-12 μm.

Inspecting the autocorrelation curve G(t) of the reference solution with the standard 3D diffusion model, the focus waist and height were calibrated. The structure factor $f$, the ratio of height to width of the focus beam, varied from day to day in the range of 6~8, and G(t) of samples was fitted using SymPhoTime (PicoQuant) software. The day to day variability was less than the variability between measurements recorded the same day. Data acquisition times were 30 s. Enzyme reaction is not expected to change the confocal volume, hence uncertainties in calibrating the confocal volume are not believed to influence the relative measurements on which this study focuses.

Fluorescence lifetime measurement. The lifetime experiments (Leica TCS SP8X, Leica, Germany) used a 100x oil immersion objective lens with numerical aperture N.A.=1.4, an excitation wavelength 488 nm with excitation at 80 MHz and a pulse width of 80 ps. Emitted fluorescence was collected using an avalanche photodiode (Micro Photon Devices, PicoQuant, Berlin, Germany) through a 500-550 nm bandpass filter and recorded using a time-correlated single-photon-counting (TCSPC) detection unit (Picoharp 300, PicoQuant) which is integrated into the microscope and saves detected photons on the fly as data is acquired. Using the microscope software (SymPhoTime, PicoQuant), this allows reconstruction of fluorescence lifetime decays.

Dynamic light scattering. A Brookhaven ZetaPALS instrument with the ZetaPlus option at 90° scattering angle and a temperature control function was used in the IBS Center for Multidimensional Carbon Materials. For DLS measurements, 30 nM dye-labeled enzymes (Urease, AChE) and the substrate solution (urea for urease, acetylthiocholine for AChE) were mixed at the desired concentration in 100 mM PBS buffer (pH 7.2) and filtered twice using 100 nm pore size syringe filter (Whatman).